# Graphene *Thickness-Graded* Transistors with Reduced Low-Frequency 1/*f* Noise

Guanxiong Liu[1], Sergey Rumyantsev[2,3], Michael Shur[2] and Alexander A. Balandin[1, *]

[1]Nano-Device Laboratory, Department of Electrical Engineering and Materials Science and Engineering Program, Bourns College of Engineering, University of California – Riverside, Riverside, California 92521 USA

[2]Center for Integrated Electronics and Department of Electrical, Computer and Systems Engineering, Rensselaer Polytechnic Institute, Troy, New York 12180 USA

[3]Ioffe Physical-Technical Institute, Russian Academy of Sciences, St. Petersburg, 194021 Russia

**Abstract**

We demonstrate graphene *thickness-graded* transistors with high electron mobility and low 1/*f* noise (*f* is a frequency). The device channel is implemented with few-layer graphene with the thickness varied from a single layer in the middle to few-layers at the source and drain contacts. It was found that such devices have electron mobility comparable to the reference single-layer graphene devices while producing lower noise levels. The metal doping of graphene and difference in the electron density of states between the single-layer and few-layer graphene cause the observed noise reduction. The results shed light on the noise origin in graphene.

High room-temperature (RT) electron mobility of graphene up to 22000 cm$^2$/Vs [1], excellent charge carrier saturation velocity of 4.5×10$^7$ cm/s [2-3], and outstanding thermal conductivity of above 3000 W/mK [4-5], exceeding that of diamond, make this material a promising candidate for the radio frequency (RF) and analog electronics applications [6-7]. Important components of the analog systems such as phase detectors have already been implemented with the triple-mode graphene transistors [8-9]. In RF and analog applications, the reduction of the low-frequency noise is important because this type of noise contributes to the phase noise of the systems as in voltage controlled oscillators or radars [10].

The low-frequency noise in graphene field-effect transistors (FETs) has the drain-current noise spectral density $S_I$~1/$f$ for the frequency $f$ below 100 kHz [11-18]. Some graphene devices also exhibit the generation-recombination (G-R) noise bulges with the time constants $\tau=1/(2\pi f_o)$ of ~0.3 – 1.1 s ($f_o$ is the corner frequency) [12]. The noise level in graphene FETs is strongly affected by the quality of the graphene-metal contacts and environmental exposure [15]. Several reports suggested that the low-frequency noise can be reduced in bilayer graphene (BLG) devices as compared to that in the single-layer graphene (SLG) devices [11, 14]. The physical mechanism for the reduction and the dominant noise sources, e.g. graphene channel vs. graphene – metal contact, are still the subjects of discussion [11-15]. However, even if BLG devices produce less noise, SLG devices demonstrate the highest charge carrier mobility, which is a major benefit for the high-frequency applications [6-7].

In this letter, we propose and demonstrate a new type of graphene devices with a graded thickness in the direction from the contacts toward the middle. Such devices combine the benefit of higher mobility of SLG and lower noise of BLG. In these devices, the main part of the channel – between the source and drain – has a thickness of one atomic plane ($n$=1), while the regions closer to the metal contacts have a thickness of two atomic planes ($n$=2) or more. We refer to this type of FETs as graphene thickness-graded (GTG) transistors. In GTG FETs, the metal contacts are made intentionally on the BLG or few layer graphene (FLG) parts avoiding any contact with the SLG channel (see Figure 1).

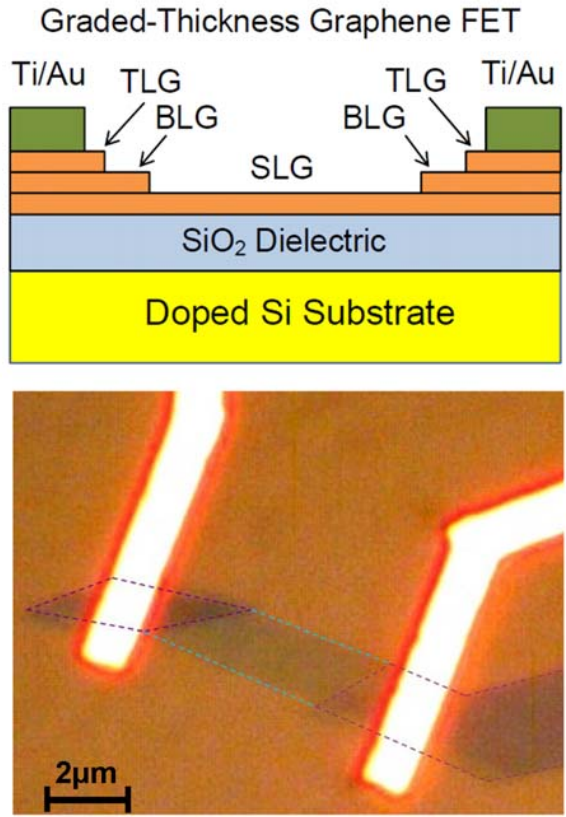

**Figure 1:** Schematic of the proposed graphene graded-thickness field-effect transistors (upper panel) and an optical microscopy image showing one of such devices (lower panel). The graphene ribbon used as the graded-thickness channel is indicated with the dash lines. The darker regions correspond to the few-layer graphene ($n$=3 in this case). The bright white bars are metal electrodes connected to the source and drain regions.

For the proof-of-concept demonstration, we produced the GTG layers by the standard exfoliation method [19] but used the flakes of the ribbon-like shape with the thickness varying from $n$=1 in the middle to $n$=3 at the both ends. Initially, the suitable GTG flakes were identified under the optical microscope. The gradation in the flake thickness was then verified with the micro-Raman spectroscopy utilizing the comparison the 2D/G peak intensity ratio and deconvolution of 2D (G') band [20-21]. All Raman spectra were measured under 633-nm laser excitation in the backscattering configuration at RT. Details of our Raman microscopy protocols were reported elsewhere [21-24]. Figure 2 shows the Raman spectra from different locations of the same GTG flake on Si/SiO$_2$ substrate. One can see the signatures of SLG in the middle, BLG in the transition region and FLG at the end of the flake [20-23].

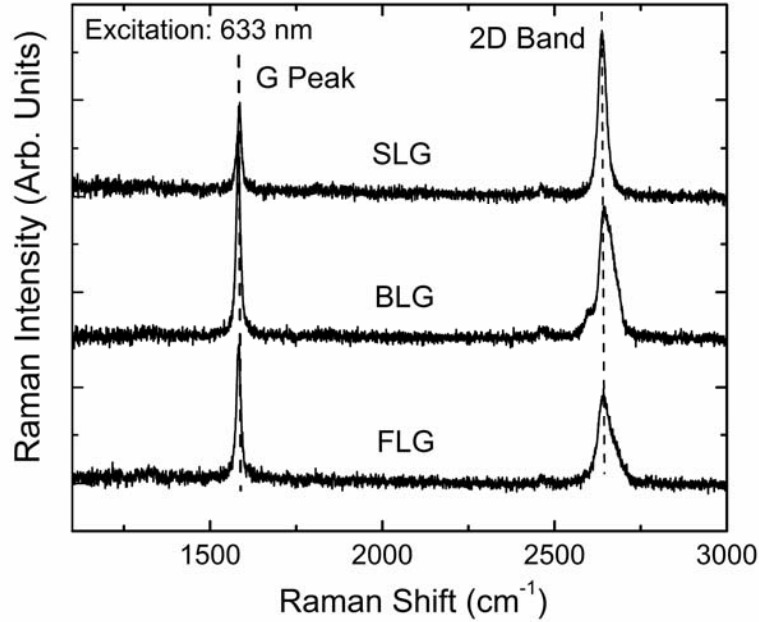

**Figure 2:** Raman spectra from different regions of the same flake used for fabrication of the channel of the graphene graded-thickness transistor. The top spectrum displaying clear signatures of the single-layer graphene was recorded from the central region of the channel. The bottom spectrum characteristic for the few-layer graphene was recorded close to the contact region.

The source and drain electrodes were fabricated by the electron beam lithography (EBL) followed by the electron beam evaporation (EBE). The electrode metals were Ti/Au with the thickness of 8-nm/80-nm, respectively. The degenerately doped p-type Si substrate acted as the back gate for tuning the electrical conductivity of the graphene channel. We have also fabricated a large number of SLG and BLG FETs (>15) to be used as references devices for comparison with the GTG FETs. An inset to Figure 3 shows the drain-source resistance $R_{DS}$ as a function of the back-gate voltage $V_{BG}$ of the GTG FET measured at ambient conditions. The $R_{DS}(V_{BG})$ dependence is similar to that of conventional SLG FETs. The fabricated devices were robust and retained their current-voltage (I-Vs) over the testing period of several weeks at ambient. The SLG, GTG and BLG FETs, fabricated using the same process, had the RT electron mobility $\mu$ values in the ranges ~5000 – 7000 cm$^2$/Vs, ~4000 – 5000 cm$^2$/Vs and ~1000 – 2000 cm$^2$/Vs, respectively. GTG FETs retained the high mobility values close to those characteristic for SLG devices. The contact resistance was estimated using the transmission line model (TLM) structures, four-probe measurements and analyzing the I-V characteristics of transistors [25]. GTG devices were characterized by the smallest contact resistance within the range 0.2-0.8 Ω-mm.

Following the I-V characterization, the low-frequency noise was measured with a spectrum analyzer (SRS 760 FFT). The device bias was applied with a "quiet" battery - potentiometer circuit. Figure 3 shows representative low-frequency noise spectra in GTG FETs. The spectra reveal 1/$f$ noise spectral density in the frequency range from 1 Hz to 100 kHz similar to that observed in SLG and BLG FETs. No G-R bulges were observed in the tested GTG FETs. We have examined the normalized noise spectral density $S_I/I^2$ dependence on the area of the device channels $S$ (here $I=I_{DS}$ is the drain-source current).

As seen in Figure 4, $S_I/I^2$ in the reference BLG devices decreases with the increasing channel area, $S$, while the 1/$f$ noise in the SLG devices shows only a weak area dependence. The strong dependence of the noise spectral density in BLG FETs on $S$ (noise level scales with the area of two-dimensional channel) indicates that the main contribution to the 1/$f$ noise comes from the graphene channel. The weak $S$ dependence in SLG FETs suggests that the contribution of the contact noise is substantial. As seen in Figure 4, GTG FETs produce less noise than SLG FETs and have the $S_I/I^2$ dependence on the channel area. This means that by using the specially designed channel, which is SLG in middle but has FLG thickness at the contact regions, we were able to reduce the metal-graphene contact contribution to the low-frequency noise. This result also provides additional evidence that the contacts in conventional SLG devices can substantially contribute to the noise level. One should note here that SLG and BGT FETs have been also investigated and compared in Ref [15]. Subsequent studies revealed that SLG FETs examined in Ref. [15], which had lower noise level, were actually GTG-type devices.

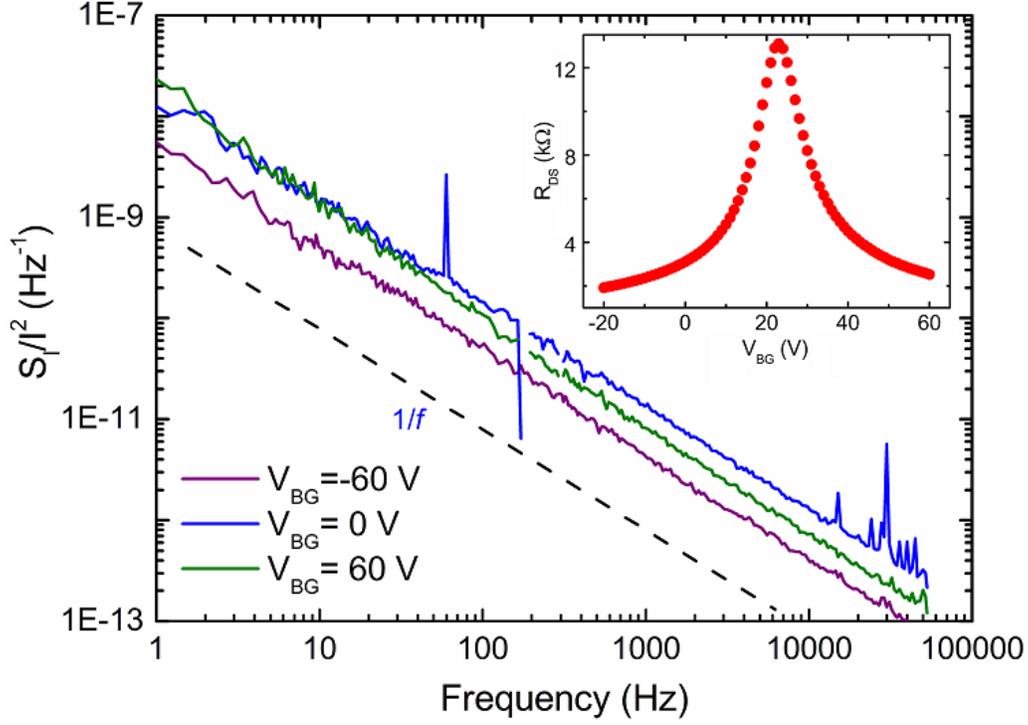

**Figure 3:** Normalized noise spectrum density as a function of frequency *f* for several values of the back-gate bias. The 1/*f* spectrum is added for comparison. The inset shows a typical drain-source resistance characteristic of the graphene thickness-graded transistor near the Dirac point.

We now offer a physical model, which explains the noise reduction in GTG FETs by the lower potential fluctuations at the metal – FLG interface as compared to those at the metal – SLG interface. Fabrication of the metal contacts to graphene leads to the metal doping of graphene via the charge transfer to reach the equilibrium conditions, and, correspondingly, results in the local shift of the Fermi level position in graphene. Theory suggests that metals with the work functions different from graphene, can dope graphene both *n*-type and *p*-type [26]. The electron density of states (DOS) in SLG in the vicinity of the charge neutrality point is low owing to the Dirac-cone linear dispersion. For this reason, even a small amount of the charge transfer from or to the metal can strongly affect the Fermi energy of graphene. The values of $\Delta E_F$=-0.23 eV and $\Delta E_F$=0.25 eV were reported for Ti and Au contacts to graphene, respectively [27]. The scanning photocurrent studies confirmed the strong non-uniform potential variations at the metal-graphene contact edge [28-29]. The quadratic energy dispersion of BLG results in DOS, which is different from that in SLG. Thus, the same amount of charge transfer between the metal and graphene – determined by the work function difference – will lead to the smaller Fermi level shifts in BLG than in SLG owing to the larger DOS in BGL (see inset to Figure 4). The potential barrier fluctuations will be smaller at the metal-BLG (or FLG) interface than in the metal-SLG interface.

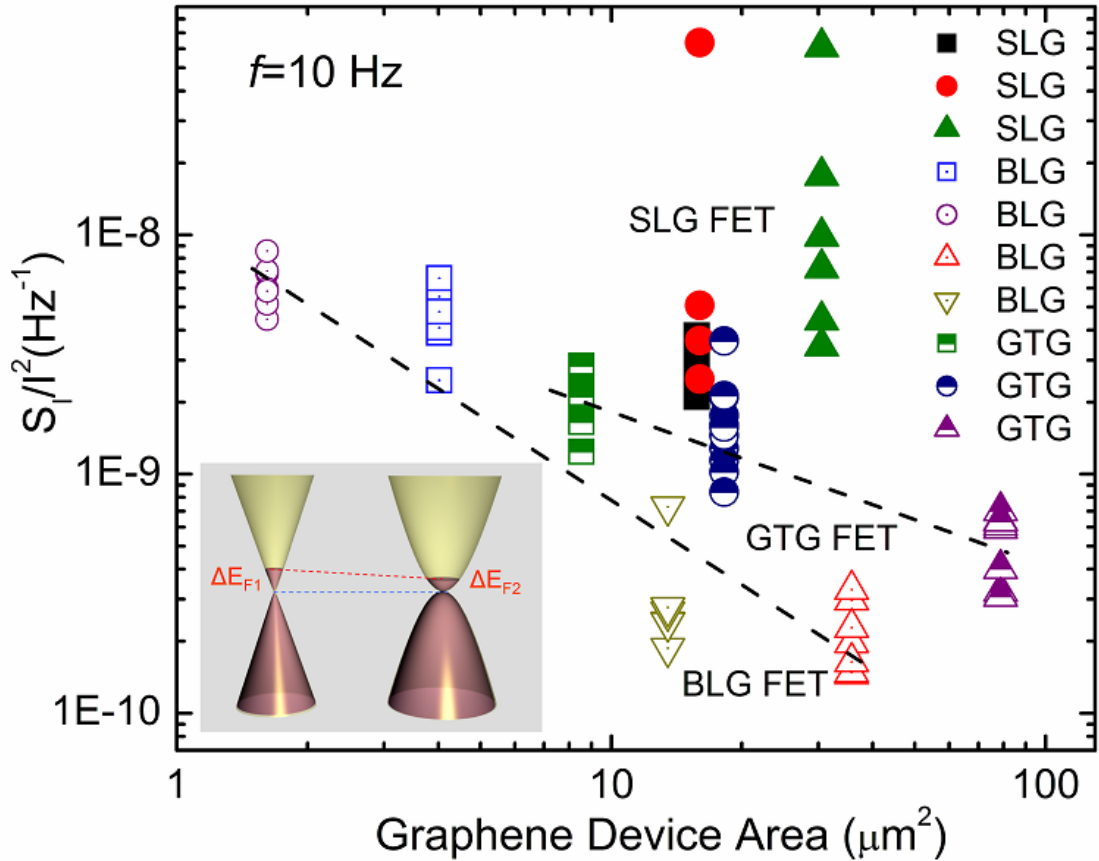

**Figure 4:** Normalized noise spectral density of the GTG FETs and the reference SLG and BLG FETs as the function of the graphene channel area. The filled symbols represent SLG, the open symbols – BLG while the half-filled symbols indicate the data-points for GTG FETs. For each device the data are shown at several biasing points within the $|V_{BG}-V_{CNP}|\leqslant 30$ V range from the charge neutrality point $V_{CNP}$. Note that GTG FETs have the reduced noise level, close to that the in BLG FETs, while revealing the electron mobilities almost as high as in SLG FETs. The inset shows the band structure of SLG and BLG in vicinity of the charge neutrality point. The same amount of the charge, transferred owing to the metal contact doping, leads the smaller local Fermi level shift in BLG devices than in SLG devices.

The potential fluctuations due to the traps at interface between Si/SiO$_2$ were identified as the origin of 1/$f$ noise in the Si metal-oxide-semiconductor field-effect transistors (MOSFETs) [30]. The local potential fluctuations can contribute to the low-frequency noise via both mobility-fluctuation and carrier number fluctuation mechanisms [10]. Owing to the discussed differences in DOS the contact between the metal and SGL will have stronger potential inhomogeneities than that between the same metal and FLG. The reason for this is not only technological but also fundamental – related to the difference in electronic band-structure between SLG and FLG. The latter explains the observed reduction of the 1/$f$ noise level in our GTG FETs. It has been previously stated that the resistivity of the metal-graphene contacts will be the performance-limiting characteristic in graphene devices [31]. The present results suggest that the metal-graphene contacts are also the important factor for the 1/$f$ noise level in graphene devices.

In conclusion, we demonstrated a new type of graphene devices – graphene thickness-graded transistors – which combine the high electron mobility of a single-layer graphene and the low 1/$f$ noise of the bilayer graphene devices. The investigation of the noise spectra in this new device revealed the contribution of the metal-graphene contact to the overall noise level and shed light on the origin of the low-frequency fluctuations in graphene devices.

**Acknowledgements**

The work at UCR was supported by the Semiconductor Research Corporation (SRC) and the Defense Advanced Research Project Agency (DARPA) through FCRP Center on Functional Engineered Nano Architectonics (FENA), the US National Science Foundation (NSF) and the US Office of Naval Research (ONR) through award N00014-10-1-0224. The work at RPI was supported by the US NSF under the auspices of I/UCRC "CONNECTION ONE" at RPI and by the NSF EAGER program. SLR acknowledges partial support from the Russian Fund for Basic Research (RFBR) grant 11-02-00013.